\documentclass{elsart}

 \usepackage{graphicx}

\begin{document}

\begin{frontmatter}
  \begin{flushleft}
  \includegraphics[width=0.3\columnwidth,height=!]{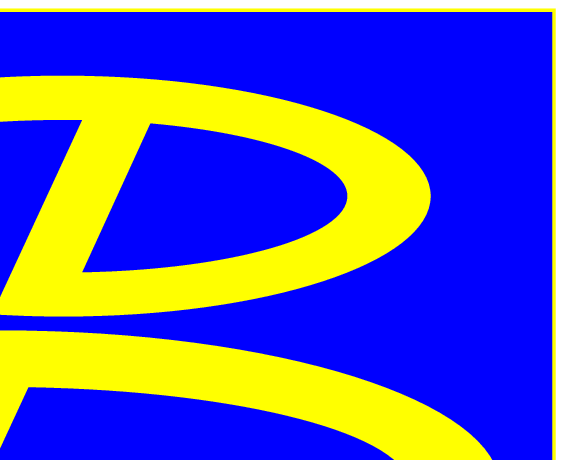}
\end{flushleft}

\begin{flushright}
\vskip -3cm
\noindent
Belle Prerpint 2002-18 \\
KEK   Preprint 2002-59\\
\end{flushright}
\vspace{2cm}

\title{Study of $B \to \rho \pi$ decays at Belle}


\collab{Belle Collaboration}
  \author[Melbourne]{A.~Gordon}, 
  \author[Taiwan]{Y.~Chao}, 
  \author[KEK]{K.~Abe}, 
  \author[TohokuGakuin]{K.~Abe}, 
  \author[TIT]{N.~Abe}, 
  \author[Niigata]{R.~Abe}, 
  \author[Tohoku]{T.~Abe}, 
  \author[Korea]{Byoung~Sup~Ahn}, 
  \author[Tokyo]{H.~Aihara}, 
  \author[Nagoya]{M.~Akatsu}, 
% ...

% ...
\address[KEK]{High Energy Accelerator Research Organization (KEK), Tsukuba, Japan}
% ...
\address[Korea]{Korea University, Seoul, South Korea}
% ...
\address[Melbourne]{University of Melbourne, Victoria, Australia}
\address[Nagoya]{Nagoya University, Nagoya, Japan}
% ...
\address[Taiwan]{National Taiwan University, Taipei, Taiwan}
% ...
\address[Niigata]{Niigata University, Niigata, Japan}
% ...
\address[TohokuGakuin]{Tohoku Gakuin University, Tagajo, Japan}
\address[Tohoku]{Tohoku University, Sendai, Japan}
\address[Tokyo]{University of Tokyo, Tokyo, Japan}
\address[TIT]{Tokyo Institute of Technology, Tokyo, Japan}
% ...
\thanks[NovaGorica]{on leave from Nova Gorica Polytechnic, Nova Gorica, Slovenia}

\begin{abstract}
  This paper describes a study of $B$ meson decays to the
  pseudoscalar-vector final state $\rho\pi$ using $31.9\times 10^6$
  $B\overline{B}$ events collected with the Belle detector at KEKB.
  The branching fractions ${\mathcal B}(B^+ \to \rho^0\pi^+) =
  (8.0^{+2.3+0.7}_{-2.0-0.7} ) \times 10^{-6}$ and ${\mathcal B}(B^0
  \to \rho^\pm \pi^\mp) = (20.8^{+6.0+2.8}_{-6.3-3.1}) \times 10^{-6}$
  are obtained.  In addition, a 90\% confidence level upper limit of
  ${\mathcal B}(B^0 \to \rho^0\pi^0) < 5.3 \times 10^{-6}$ is reported.
\end{abstract}

\end{frontmatter}

The decays of $B$-mesons to pseudoscalar and vector particles provide
opportunities for investigating the phenomenon of $CP$ violation. Of
particular interest are the quasi-two-body $B \to \rho \pi$ decays to
three-pion final states. These can be used to measure the angles
$\phi_2$ and $\phi_3$ of the unitarity triangle: $\phi_2$ can be
determined from a full Dalitz plot analysis of the modes $B^0 \to
\rho^\pm \pi^\mp$ and $B^0 \to \rho^0 \pi^0 $~\cite{pipipi-phi2};
$\phi_3$ can be extracted from the interference of $B^+ \to
\chi_{\mbox{\tiny c0}} \pi^+$ and $B^+ \to \rho^0 \pi^+$ in $B^+ \to
\pi^+ \pi^- \pi^+$ decays~\cite{pipipi-phi3}. While the data set used
here is too small for this type of analysis, it can be used to study
the branching fractions of these decays. The ratio of branching
fractions $R = {\mathcal B}(B^0 \to \rho^\pm \pi^\mp) / {\mathcal
  B}(B^+ \to \rho^0 \pi^+)$ is particularly interesting as theoretical
predictions for the value of $R$ vary over a wide range depending on
the assumptions made in its
calculation~\cite{ratio-bauer,ratio-deandrea,chen,lu,tandean,gardner}.

In this paper, a search for $B$ meson decays of the type $B \to \rho
\pi$ is described.  Both the neutral ($B^0 \to \rho^\pm \pi^\mp$ and
$B^0 \to \rho^0 \pi^0$) and charged ($B^+ \to \rho^0 \pi^+$) modes are
examined. Here and throughout the text, inclusion of charge conjugate
modes is implied and for the neutral decay, $B^0 \to \rho^\pm
\pi^\mp$, the notation implies a sum over both the modes. The data
sample used in this analysis was taken by the Belle
detector~\cite{Belle} at KEKB~\cite{KEKB}, an asymmetric storage ring
that collides 8 GeV electrons against 3.5 GeV positrons. This produces
$\Upsilon(4S)$ mesons that decay into $B^0\overline{B^0}$ or $B^+B^-$
pairs. The data sample has an integrated luminosity of 29.4 fb$^{-1}$
and consists of $31.9 \times 10^6 ~B\overline{B}$ pairs.

The Belle detector is a general purpose spectrometer based on a $1.5$
T superconducting solenoid magnet.  Charged particle tracking is
achieved with a three-layer double-sided silicon vertex detector
(SVD) surrounded by a central drift chamber (CDC) that consists of
$50$ layers segmented into 6 axial and 5 stereo super-layers. The CDC
covers the polar angle range between $17^\circ$ and $150^\circ$ in the
laboratory frame, which corresponds to 92\% of the full centre of mass
(CM) frame solid angle. Together with the SVD, a transverse momentum
resolution of $(\sigma_{p_t}/p_t)^2 = (0.0019 \,p_t)^2 + (0.0030)^2$
is achieved, where $p_t$ is in GeV/$c$.

Charged hadron identification is provided by a combination of three
devices: a system of $1188$ aerogel \v{C}erenkov counters (ACC)
covering the momentum range $1$--$3.5$ GeV/$c$, a time-of-flight
scintillation counter system (TOF) for track momenta below $1.5$
GeV/$c$, and $dE/dx$ information from the CDC for particles with very
low or high momenta. Information from these three devices is combined
to give the likelihood of a particle being a kaon, $L_{K}$, or pion,
$L_{\pi}$. Kaon-pion separation is then accomplished based on the
likelihood ratio $L_{\pi}$/($L_{\pi}+L_{K})$. Particles with a
likelihood ratio greater than $0.6$ are identified as pions. The pion
identification efficiencies are measured using a high momentum
$D^{*+}$ data sample, where $D^{*+}\to D^0\pi^+$ and $D^0\to
K^-\pi^+$. With this pion selection criterion, the typical efficiency
for identifying pions in the momentum region $0.5$ GeV/$c$ $<p<4$
GeV/$c$ is $(88.5 \pm 0.1)$\%. By comparing the $D^{*+}$ data sample
with a Monte Carlo (MC) sample, the systematic error in the particle
identification (PID) is estimated to be 1.4\% for the mode with three
charged tracks and 0.9\% for the modes with two.

Surrounding the charged PID devices is an electromagnetic calorimeter
(ECL) consisting of $8736$ CsI(Tl) crystals with a typical
cross-section of $5.5 \times 5.5 $ cm$^2$ at the front surface and
$16.2$ $X_{0}$ in depth. The ECL provides a photon energy resolution
of $(\sigma_E/E)^2 = 0.013^2 + (0.0007/E)^2 + (0.008/E^{1/4})^2$,
where $E$ is in GeV.

Electron identification is achieved by using a combination of $dE/dx$
measurements in the CDC, the response of the ACC and the position and
shape of the electromagnetic shower from the ECL. Further information is
obtained from the ratio of the total energy registered in the
calorimeter to the particle momentum, $E/p_{lab}$.

Charged tracks are required to come from the interaction point and
have transverse momenta above $100$ MeV/$c$.
Tracks consistent with being an electron are rejected and the
remaining tracks must satisfy the pion identification requirement. The
performance of the charged track reconstruction is studied using high
momentum $\eta \to \gamma\gamma$ and $\eta \to \pi^+\pi^-\pi^0$
decays. Based on the relative yields between data and MC, we assign a
systematic error of $2\%$ to the single track reconstruction
efficiency.

Neutral pion candidates are detected with the ECL via their decay
$\pi^0 \to \gamma \gamma$. The $\pi^0$ mass resolution, which is
asymmetric and varies slowly with the $\pi^0$ energy, averages to
$\sigma = 4.9$ MeV/$c^2$. The neutral pion candidates are selected
from $\gamma \gamma$ pairs by requiring that their invariant mass to
be within $3\sigma$ of the nominal $\pi^0$ mass.


To reduce combinatorial background, a selection criteria is applied to
the photon energies and the $\pi^0$ momenta. Photons in the barrel
region are required to have energies over 50 MeV, while a 100 MeV
requirement is made for photons in the end-cap region. The $\pi^0$
candidates are required to have a momentum greater than 200 MeV/$c$ in
the laboratory frame. For $\pi^0$s from $B\overline{B}$ events, the
efficiency of these requirements is about 40\%. Furthermore, the
$\pi^0$ helicity, defined as the absolute cosine of the angle between
the photon direction in the $\pi^0$ rest frame and the $\pi^0$
direction in the lab frame, is required to be less than 0.95. The
behavior of the $\pi^0$ reconstruction efficiency as a function of
momentum is studied using $D^{*+}\to D^0\pi^+$ decays by comparing the
relative yields between data and MC in the $D^0\to K^-\pi^+\pi^0$ and
$K^-\pi^+$ sub-channels. The residual systematic uncertainties are
$9.8\%$ for low momentum $\pi^0$s ($p<500$ MeV/$c$) and $7.7\%$
otherwise.


Neutral $\rho$ candidates are reconstructed using the decay $\rho^0
\to \pi^+ \pi^-$ with the invariant mass of the charged pion pair
required to be between 0.6 GeV/$c^2$ and 0.95 GeV/$c^2$. The decay
$\rho^+ \to \pi^0\pi^+$ is used to reconstruct charged $\rho$
candidates with the invariant mass of the pion pair required to be
between 0.62 GeV/$c^2$ and 0.92 GeV/$c^2$.


Candidate $B$ mesons are identified by pairing $\rho$ and $\pi$
candidates that pass the above criteria. The candidates are then
selected using the beam-constrained mass $M_{bc} =
\sqrt{E^2_{\mbox{\scriptsize beam}} - p_B^2}$ and the energy
difference $\Delta E = E_B - E_{\mbox{\scriptsize beam}}$. Here,
$p_{B}$ and $E_B$ are the momentum and energy of a $B$ candidate in
the CM frame and $E_{\mbox{\scriptsize beam}}$ is the CM beam energy.
An incorrect mass hypothesis for a pion or kaon produces a shift of
about 46 MeV in $\Delta E$, providing extra discrimination between
these particles. The width of the $M_{bc}$ distributions is primarily
due to the beam energy spread and is well modelled with a Gaussian of
width 3.3 MeV/$c^2$ for the modes with a neutral pion and 2.7
MeV/$c^2$ for the mode without. The $\Delta E$ distribution is found
to be asymmetric with a small tail on the lower side for the modes
with a $\pi^0$. This is due to $\gamma$ interactions with material in
front of the calorimeter and shower leakage out of the calorimeter.
The $\Delta E$ distribution can be well modelled with a Gaussian when
no neutral particles are present. Events with $5.2$ GeV/$c^2$ $<
M_{bc} < 5.3$ GeV/$c^2$ and $|\Delta E |< 0.3 $ GeV are selected for
the final analysis.

The dominant background comes from continuum $e^+e^-\rightarrow
q\overline{q}$ ($q$ = $u$, $d$, $s$, $c$) production. This background
is suppressed using variables that quantify the difference in the
event topology between spherical $B\overline{B}$ events and jet-like
$q\overline{q}$ events. The most powerful suppression is achieved
using variables derived from the Fox-Wolfram moments~\cite{FW} but
altered to contain the additional information of whether or not a
track comes from the $B$ candidate in the event~\cite{SFW}. These
modified moments enhance the discriminating power of the original
Fox-Wolfram moments and are defined as

\begin{displaymath}
 R_l = { \sum_{i,k} |p_i||p_k|P_l(\cos\theta_{ik}) \over
               \sum_{i,k} |p_i||p_k|}
{,~~}
  r_l = {\sum_{i,j} |p_i||p_j|P_l(\cos\theta_{ij}) \over
               \sum_{i,j} |p_i||p_j|},
\end{displaymath}
where $p$ indicates the particle momentum, $l$ runs from 1 to 4 and
$P_l$ is the Legendre polynomial of $l$th order, $k$ runs over the
daughters of a $B$ candidate and $i$ and $j$ enumerate photons and
charged particles in the event not associated with the $B$ candidate.
Since $R_1$, $R_3$ and $r_1$ are found to be correlated with $M_{bc}$,
they are not used. Another event topology variable used is
$S_\perp$, defined as the scalar sum of the transverse momenta (with
respect to the thrust axis) of the particles outside a $45^{\circ}$
cone around the thrust axis, divided by the scalar sum of the momenta
of all the particles. These variables then are combined to form a
Fisher discriminant~\cite{fisher}
\begin{displaymath}
{\mathcal F} = \sum_{l=2,4} \alpha_l R_l +  \sum_{l=2,3,4} \beta_l r_l +
\gamma S_\perp,
\end{displaymath}
where $\alpha_l$, $\beta_l$ and $\gamma$ are optimized to maximize the
separation between signal and continuum events. Also used for
continuum suppression is the $B$ flight direction, $\cos\theta_B$,
defined as the cosine of the angle the $B$ meson momentum vector makes
with the positron beam axis. As $\cos\theta_B$ is uncorrelated with
the other continuum suppression variables, it is left out of $\mathcal F$.

Probability density functions (PDFs) are obtained for ${\mathcal F}$ and
$\cos\theta_B$ using MC simulations for signal and background. The
PDFs are then combined to form a likelihood ratio ${\mathcal L} = {\mathcal
  L}_s/({\mathcal L}_s + {\mathcal L}_{q\overline{q}})$, where ${\mathcal L}_{s}$
and ${\mathcal L}_{q\overline{q}}$ are the product of the signal and
background PDFs, respectively. Signal events are selected by requiring
that ${\mathcal L}>0.9$, which has a signal efficiency of about 40\% for
modes with a $\pi^0$ and 35\% for the all charged particle mode. The
systematic error in the likelihood ratio requirement is determined
using a sample of $B^+ \to {\overline D^0} \pi^+$ decays. By comparing
the yields in data and MC after the likelihood ratio requirement, the
systematic errors are determined to be 4\% for the modes with a
$\pi^0$ and 6\% for the mode without.

The final variable used for continuum suppression is the $\rho$
helicity angle, $\theta_h$, defined as the angle between the direction
of the decay pion from the $\rho$ in the $\rho$ rest frame and the
$\rho$ in the $B$ rest frame. The requirement of $|\cos\theta_h |>0.3$
is made independently of the likelihood ratio as it is effective in
suppressing the background from $B$ decays as well as the $q
\overline{q}$ continuum.

Although continuum events are the largest background, the mode with
all charged particles has other backgrounds in the signal region from
$B$ meson decays. To study this, a large MC sample of $B\overline{B}$
events is used~\cite{gen-b}. The largest component of this background
is found to come from decays of the type $B \to {D \pi}$; when the $D$
meson decays via $D \to \pi^+ \pi^-$, events can directly reach the
signal region while the decay $D \to K^- \pi^+$ can reach the signal
region with the kaon misidentified as a pion. Decays with $J/\psi$ and
$\psi(2S)$ mesons can also populate the signal region if both the
daughter leptons are misidentified as pions. These events are excluded
by making requirements on the invariant mass of the intermediate
particles: $|M(\pi^+\pi^-) - M_{D^0}|> 0.14$ GeV/$c^2$,
$|M(\pi^+\pi^0) - M_{D^+}|> 0.05$ GeV/$c^2$, $|M(\pi^+\pi^-) -
M_{J/\psi}|> 0.07$ GeV/$c^2$ and $|M(\pi^+\pi^-) - M_{\psi(2S)}|>
0.05$ GeV/$c^2$. The widest cut is made around the $D^0$ mass to
account for the mass shift due to misidentifying the kaons in $D^0$
decays as pions.

\begin{figure}[!htb]
  \begin{center}                 
  \includegraphics*[width=12cm]{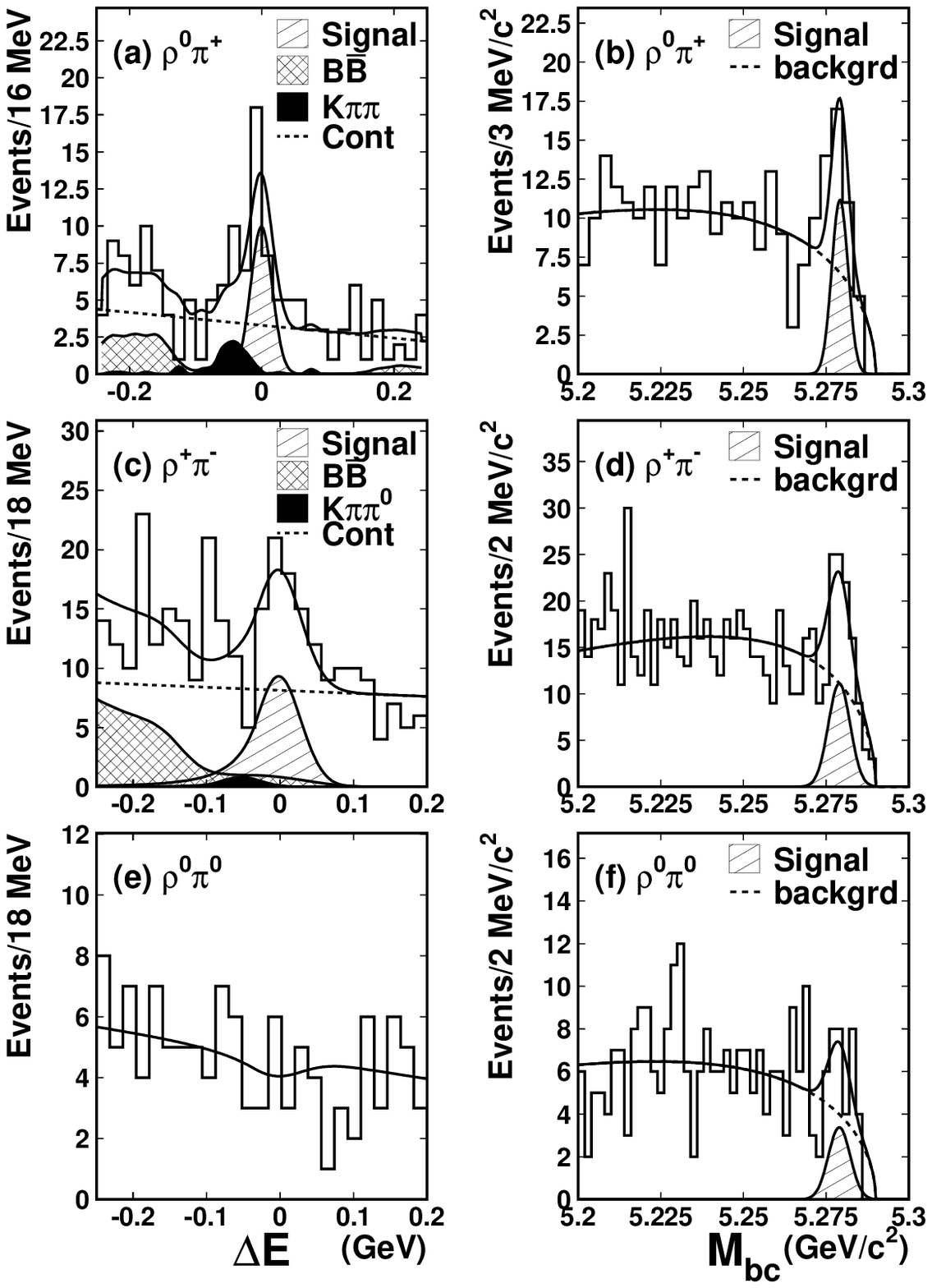} 
  \end{center}

   \caption{ The $\Delta E$ (left) and $M_{bc}$ (right) fits to the three
     $B \to \rho \pi$ modes: $\rho^0 \pi^+$, $\rho^+ \pi^-$ and
     $\rho^0 \pi^0$. The histograms show the data, the solid lines
     show the total fit and the dashed lines show the continuum
     component. In (a) the contribution from the $B \to \rho \rho$ and
     $B \to hh$ modes is shown by the cross hatched component. In (c)
     the cross hatched component shows the contribution from the $b
     \to c$ transition and $B \to \rho \rho$ modes. }
  \label{fig:mb-de-fits}

\end{figure}

Fig.~\ref{fig:mb-de-fits} shows the $\Delta E$ and $M_{bc}$
distributions for the three modes analysed after all the selection
criteria have been applied. The $\Delta E$ and $M_{bc}$ plots are
shown for events that lie within 3$\sigma$ of the nominal $M_{bc}$ and
$\Delta E$ values, respectively. The signal yields are obtained by
performing maximum likelihood fits, each using a single signal
function and one or more background functions.  The signal functions
are obtained from the MC and adjusted based on comparisons of $B^+ \to
\overline{D^0} \pi^+$ decays in the data and MC. All parameters in the
signal function are fixed except the overall normalization, which is
allowed to float. The final branching fractions are extracted using
the $\Delta E$ signal yield, as this provides a kinematic separation
of the background modes. In calculating the branching fractions, the
decay rates of the $\Upsilon(4S)$ into $B^+B^-$ and
$B^0\overline{B^0}$ are assumed to be equal.

The $M_{bc}$ distribution for all modes is fitted with a single
Gaussian and an ARGUS background function~\cite{ARGUS}. The
normalization of the ARGUS function is left to float and shape of the
function is fixed from the $\Delta E$ sideband: $-0.25$ GeV $<\Delta
E< -0.08$ GeV and $5.2$ GeV/$c^2$ $< M_{bc} < 5.3$ GeV/$c^2$. For the
mode with only charged pions in the final state, the $\Delta E$
distribution is fitted with a single Gaussian for the signal and a
linear function with fixed shape for the continuum background. The
normalization of the linear function is left to float and the slope is
fixed from the $M_{bc}$ sideband, $5.2$ GeV/$c^2$ $< M_{bc} < 5.26 $
GeV/$c^2$, $|\Delta E |< 0.3 $ GeV. There are also other rare $B$
decays that are expected to contaminate the $\Delta E$ distribution.
For the mode without a $\pi^0$, these modes are of the type $B^0 \to
h^+h^-$ (where $h$ denotes a $\pi$ or $K$), $B \to \rho \rho$
(including all combinations of charged and neutral $\rho$ mesons,
where the polarizations of the $\rho$ mesons are assumed to be
longitudinal) and $B \to K\pi\pi$ (including the decays $B^+ \to
\rho^0 K^+$, $B^+ \to K^{*0} \pi^+ $, $B^+ \to K_0^*(1430)^0 \pi^+ $,
$B^+ \to f_0(980) K^+$ and $B^+ \to f_0(1370) K^+$)~\cite{khh}. These
background modes are accounted for by using smoothed histograms whose
shapes have been determined by combining MC distributions. The three
$B \to \rho \rho$ modes are combined into one histogram. The
normalization of this component is allowed to float in the fit due to
the uncertainty in the branching fractions of the $B \to \rho \rho$
modes. Likewise, the $B \to hh$ and all the $B \to K\pi\pi $ modes are
combined to form one $hh$ and one $ K\pi\pi$ component. The
normalizations of these components are fixed to their expected yields,
which are calculated using efficiencies determined by MC and branching
fractions measured by previous Belle analyses~\cite{khh,hh}.

The $\Delta E$ fits for the modes with a $\pi^0$ in the final state
have the signal component modelled by a Crystal Ball
function~\cite{crystal-ball} to account for the asymmetry in the
$\Delta E$ distribution. As for the $B^+ \to \rho^0 \pi^+$ mode, the
continuum background is modelled by a linear function with fixed
slope. Unlike the $B^+ \to \rho^0 \pi^+$ mode, a component is included
for the background from the $b\to c$ transition. The parameterization
for rare $B$ decays includes one component for the $B \to K\pi\pi^0$
modes ($B^0 \to \rho^+ K^-$ and $B^0 \to K^{*+}
\pi^-$)~\cite{khpi_zero} and one for all the $B \to \rho \rho$ modes.
The normalization of the $B \to \rho \rho$ component is left to float
while the other components from $B$ decays are fixed to their expected
yields.

Table~\ref{tab:results} summarizes the results of the $\Delta E$ fits,
showing the number of events, signal yields, reconstruction
efficiencies, statistical significance and branching fractions or
upper limits for each fit. The statistical significance is defined as
$\sqrt{-2 \ln({\mathcal L}_0/{\mathcal L}_{max})}$, where ${\mathcal
  L}_{max}$ denotes the likelihood at the nominal signal yield and
${\mathcal L}_0$ is the likelihood with the signal yield fixed to
zero. For the $\rho^0\pi^+$ and $\rho^+ \pi^-$ modes, the yields
obtained from the $\Delta E$ fit are $24.3^{+6.9}_{-6.2}$ and
$44.6^{+12.8}_{-13.4}$, respectively. Both signals have a significance
of over 3$\sigma$ and consistent signal yields are obtained from the
$M_{bc}$ distributions. No excess above background is seen in the
$\Delta E$ fit for the $\rho^0\pi^0$ mode and an excess of $14 \pm 6$
events is seen in the $M_{bc}$ distribution. The $M_{bc}$ yield has a
significance of $2.3 \sigma$ and may be due to background from rare
$B$ decays. An upper limit (UL) is obtained for this mode based on the
$\Delta E$ yield. The upper limit is calculated using the 90\%
confidence level UL of the signal yield ($N_s$), obtained by
integrating the likelihood function to be $ \int^{UL}_0 {{\mathcal
    L}(N_s) dN_s } / \int^{\infty}_0 {{\mathcal L}(N_s) dN_s} = 0.9$,
where ${\mathcal L}(N_s)$ denotes the maximum likelihood with the
signal yield fixed at $N_s$.

The systematic errors in the branching fractions are obtained by
quadratically summing the systematic uncertainties in the signal
yield, tracking, particle identification, $\pi^0$ reconstruction and
the likelihood ratio requirement. The systematic error in the fitted
signal yield is estimated by independently varying each fixed
parameter in the fit by $1\sigma$. The final results are ${\mathcal B}(B^+
\to \rho^0\pi^+) = ( 8.0^{+2.3+0.7}_{-2.0-0.7} ) \times 10^{-6}$ and
${\mathcal B}(B^0 \to \rho^\pm \pi^\mp) = (20.8^{+6.0+2.8}_{-6.3-3.1}
)\times 10^{-6}$ where the first error is statistical and the second
is systematic. For the $\rho^0\pi^0$ mode, one standard deviation of
the systematic error is added to the statistical limit to obtain a
conservative upper limit at 90\% confidence of $5.3 \times 10^{-6}$.

\begin{figure}[!htb]
  \begin{center}
    \includegraphics*[width=15cm]{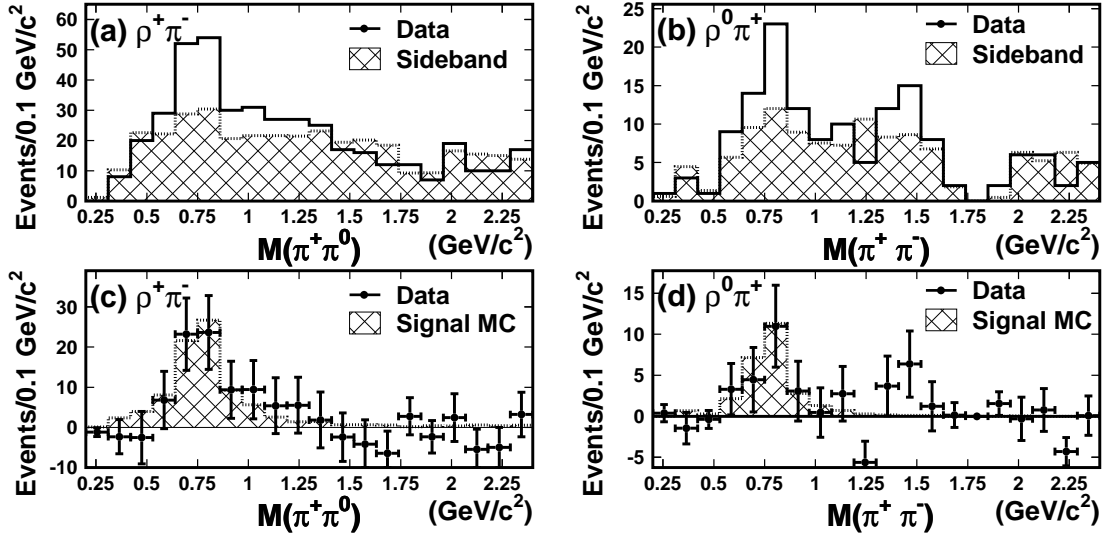}
  \end{center}
  \caption{ The $M(\pi\pi)$ distributions for $B^0 \to \rho^\pm \pi^\mp$
    (left) and $B^+\to \rho^0\pi^+$ (right) events in the signal
    region. Plots (a) and (b) show sideband events superimposed; plots
    (c) and (d) show the sideband subtracted plots with signal MC
    superimposed.}
  
  \label{fig:pp-imp}
\end{figure}

\begin{figure}[!htb]
  \begin{center}
    \includegraphics*[width=15cm]{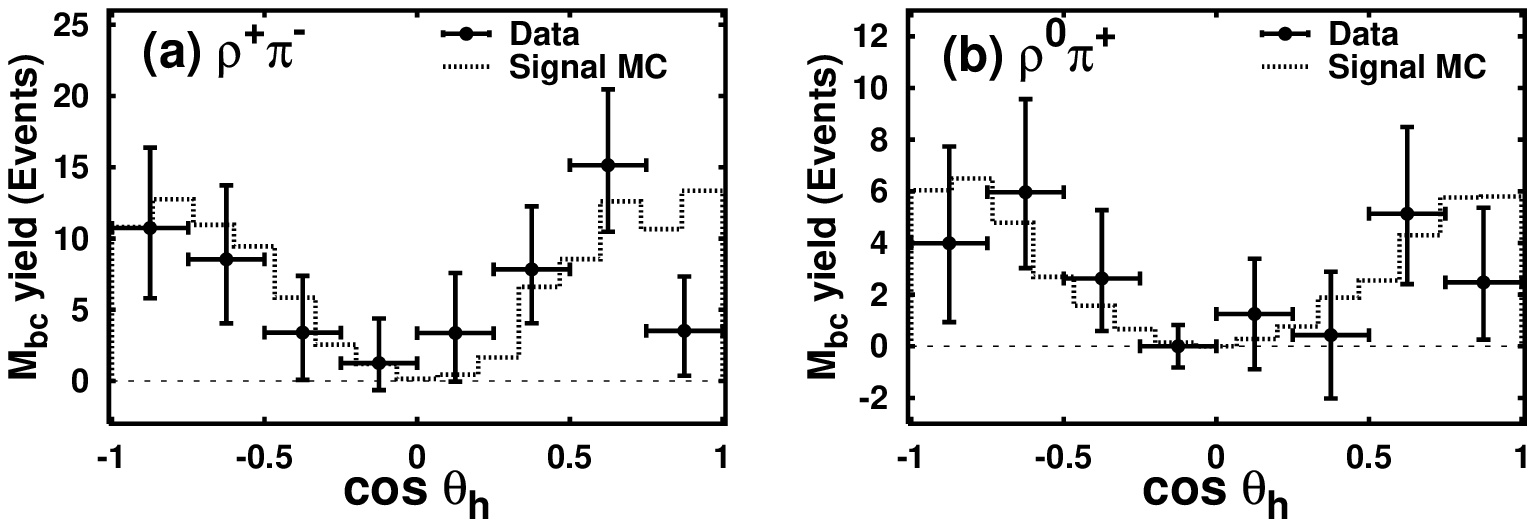}
  \end{center}
  \caption{ The $\rho$ meson helicity distributions for
    $B^0 \to \rho^\pm \pi^\mp$ (a) and $B^+\to \rho^0\pi^+$ (b).
    Signal MC distributions are shown superimposed.}
  \label{fig:helicity}
\end{figure}

The possibility of a nonresonant $B \to \pi \pi \pi$ background is
also examined. To check for this type of background, the $M_{bc}$ and
$\Delta E$ yields are determined for different $\pi\pi$ invariant mass
bins. By fitting the $M_{bc}$ distribution in $\pi\pi$ invariant mass
bins with $B \to \rho \pi$ and $B \to \pi \pi \pi$ MC distributions,
the nonresonant contribution is found to be below 4\%. To account for
this possible background, errors 3.7\% and 3.2\% are added in
quadrature to the systematic errors of the $\rho^+ \pi^-$ and $\rho^0
\pi^+$ modes, respectively.
The $\pi\pi$ invariant mass distributions are shown in
Fig.~\ref{fig:pp-imp}. Two plots are shown for the $\rho^+ \pi^-$ and
$\rho^0 \pi^+$ modes, one with events from the $M_{bc}$ sideband
superimposed over the events from the signal region (upper) and one
with events from signal MC superimposed over events from the signal
region with the sideband subtracted (lower). Fig.~\ref{fig:helicity}
shows the distribution of the helicity variable, $\cos\theta_h$, for
the two modes with all selection criteria applied except the helicity
condition. Events from $\rho \pi$ decays are expected to follow a
$\cos^2\theta$ distribution while nonresonant and other background
decays have an approximately uniform distribution. The helicity plots
are obtained by fitting the $M_{bc}$ distribution in eight helicity
bins ranging from $-1$ to $1$. The $M_{bc}$ yield is then plotted
against the helicity bin for each mode and the expected MC signal
distributions are superimposed. Both the $\pi\pi$ mass spectrum and
the helicity distributions provide evidence that the signal events are
consistent with being from $\rho\pi$ decays.

The results obtained here can be used to calculate the ratio of
branching fractions $R = {\mathcal B} (B^0 \to \rho^\pm \pi^\mp) / {\mathcal
  B} (B^+ \to \rho^0 \pi^+)$, which gives $R = 2.6 \pm 1.0 \pm 0.4$,
where the first error is statistical and second is systematic. This is
consistent with values obtained by CLEO ~\cite{CLEO-res} and BaBar
~\cite{Babar-rho+pi-,Babar-rho0pi+} as shown in Table~\ref{tab:r}.
Theoretical calculations done at tree level assuming the factorization
approximation for the hadronic matrix elements give
$R\sim6$~\cite{ratio-bauer}.
Calculations that include penguin contributions, off-shell $B^*$
excited states or additional $\pi\pi$
resonances~\cite{ratio-deandrea,chen,lu,tandean,gardner} might yield better
agreement with the the measured value of $R$.


In conclusion, statistically significant signals have been observed in
the $B \to \rho \pi$ modes using a $31.9 \times 10^6 ~B\overline{B}$ event
data sample collected with the Belle detector. The branching fractions
obtained are ${\mathcal B}(B^+ \to \rho^0\pi^+) = (
8.0^{+2.3+0.7}_{-2.0-0.7}) \times 10^{-6}$ and ${\mathcal B}(B^0 \to
\rho^\pm \pi^\mp) = (20.8^{+6.0+2.8}_{-6.3-3.1}) \times 10^{-6}$ where
the first error is statistical and the second is systematic. No
evidence was seen for the mode $B^0 \to \rho^0 \pi^0$ and an upper
limit of $5.3 \times 10^{-6}$ at 90\% confidence level was obtained.


We wish to thank the KEKB accelerator group for the excellent
operation of the KEKB accelerator. We acknowledge support from the
Ministry of Education, Culture, Sports, Science, and Technology of
Japan and the Japan Society for the Promotion of Science; the
Australian Research Council and the Australian Department of Industry,
Science and Resources; the National Science Foundation of China under
contract No.~10175071; the Department of Science and Technology of
India; the BK21 program of the Ministry of Education of Korea and the
CHEP SRC program of the Korea Science and Engineering Foundation; the
Polish State Committee for Scientific Research under contract
No.~2P03B 17017; the Ministry of Science and Technology of the Russian
Federation; the Ministry of Education, Science and Sport of the
Republic of Slovenia; the National Science Council and the Ministry of
Education of Taiwan; and the U.S.\ Department of Energy.



\begin{table}[!htb]
\caption{ $\Delta E$ fit results. Shown for each mode are the number of
  events in the fit, the signal yield, the reconstruction efficiency, the
  branching fraction ($\mathcal B$) or 90\% confidence level  upper limit (UL)
  and the statistical significance of the fit. The first error in the
  branching  fraction is statistical, the second is systematic.}
\begin{center}
\begin{tabular}{cccccc} \hline
Channel & Events &   Signal Yield &Eff.(\%) & $\mathcal B$/UL( $\times 10^{-6}$) & Significance \\  \hline

$\rho^0\pi^+$ & 154 & $24.3^{+6.9}_{-6.2}$    & 9.6 &
$8.0^{+2.3+0.7}_{-2.0-0.7}$ & $4.4\sigma$ \\

$\rho^+\pi^-$ & 301  & $44.6^{+12.8}_{-13.4}$ & 6.8 & 
$20.8^{+6.0+2.8}_{-6.3-3.1}$ & $3.7\sigma$ \\

$\rho^0\pi^0$ & 116 & $-4.4\pm8.5$ &  8.5 & $<5.3$ & - \\ \hline
\end{tabular}
\end{center}


\label{tab:results}

\end{table}

\begin{table}[!htb]

  \caption{ A comparison between the results of the Belle, BaBar and CLEO
  experiments for the $B \to \rho \pi$ branching fractions and the ratio
  $R = {\mathcal B}(B^0 \to \rho^\pm \pi^\mp) / {\mathcal B}(B^+ \to \rho^0
  \pi^+)$. The CLEO results come from reference~\protect\cite{CLEO-res}. The
  BaBar results for $\rho^+\pi^-$ and $\rho^0\pi^+$ come from references
  \protect\cite{Babar-rho+pi-} and \protect\cite{Babar-rho0pi+}, respectively.}

\begin{center}
\begin{tabular}{cccc} \hline
Experiment & ${\mathcal B}(B^0 \to \rho^\pm \pi^\mp)\times 10^{-6}$ & ${\mathcal B}(B^+
  \to \rho^0\pi^+)\times 10^{-6}$ & $R$  \\ \hline

  Belle & $20.8^{+6.0+2.8}_{-6.3-3.1}$  & $8.0^{+2.3+0.7}_{-2.0-0.7}$
  & $2.6\pm1.1$ \\

  BaBar  & $28.9 \pm 5.4 \pm 4.3 $ &  $24 \pm 8 \pm 3$ &$1.2 \pm 0.5$\\

  CLEO & $27.6^{+8.4}_{-7.4} \pm 4.2 $ & $10.4^{+3.3}_{-3.4}\pm 2.1 $ & $2.7 \pm 1.3$ \\ \hline 
\end{tabular}
\end{center}
\label{tab:r}
\end{table}

\end{document}